\documentclass{article}

\PassOptionsToPackage{numbers, compress}{natbib}

\usepackage[sglblindworkshop, final]{neurips_2025}

\usepackage{amsmath,amssymb,amsfonts}
\usepackage{algorithmic}
\usepackage{graphicx}
\usepackage{textcomp}
\usepackage{xcolor}
\usepackage{multirow}
\usepackage[table]{xcolor}
\usepackage{amsmath}
\usepackage{graphicx}
\usepackage{booktabs}

\newcommand{\red}[1]{\textcolor{red}{\scriptsize{$\downarrow$#1}}}
\newcommand{\green}[1]{\textcolor{green!60!black}{\scriptsize{$\uparrow$#1}}}
\newcommand{\neutral}[1]{\textcolor{gray}{\scriptsize{#1}}}

\newcommand{\ppagreen}[1]{\textcolor{green!40!black}{#1$\downarrow$}}
\newcommand{\ppared}[1]{\textcolor{red!30!black}{#1$\uparrow$}}

\def\BibTeX{{\rm B\kern-.05em{\sc i\kern-.025em b}\kern-.08em
    T\kern-.1667em\lower.7ex\hbox{E}\kern-.125emX}}

\workshoptitle{ML for Systems}

\usepackage[utf8]{inputenc} 
\usepackage[T1]{fontenc}    
\usepackage{hyperref}       
\usepackage{url}            
\usepackage{booktabs}       
\usepackage{amsfonts}       
\usepackage{nicefrac}       
\usepackage{microtype}      
\usepackage{xcolor}         

\title{Automated Multi-Agent Workflows for RTL Design}

\author{%
  Amulya Bhattaram\thanks{Equal contribution; alphabetical order. $^\dagger$Corresponding author.}, Janani Ramamoorthy\footnotemark[1], Ranit Gupta, \\\textbf{Diana Marculescu, Dimitrios Stamoulis$^\dagger$} \\
  Chandra Family Department of Electrical and Computer Engineering\\
  The University of Texas at Austin\\
  Austin, TX, USA 78712 \\
  \texttt{\{abhattaram, janani.ram, ranitgupta, dianam, dstamoulis\}@utexas.edu} \\
}

\begin{document}

\maketitle

\begin{abstract}
The rise of agentic AI workflows unlocks novel opportunities for computer systems design and optimization. However, for specialized domains such as program synthesis, the relative scarcity of HDL and proprietary EDA resources online compared to more common programming tasks introduces challenges, often necessitating task-specific fine-tuning, high inference costs, and manually-crafted agent orchestration. In this work, we present \textit{VeriMaAS}, a \textit{multi-agent} framework designed to automatically compose agentic workflows for RTL code generation. Our key insight is to integrate formal verification feedback from HDL tools directly into workflow generation, reducing the cost of gradient-based updates or prolonged reasoning traces. Our method improves synthesis performance by 5–7\% for pass@k over fine-tuned baselines, while requiring only a few hundred ``training'' examples, representing an order-of-magnitude reduction in supervision cost.
\end{abstract}

\section{Introduction}
\label{sec:intro}
Agentic AI presents exciting opportunities for system optimization and design~\cite{kwon2023efficient, fore2024geckopt, xu2025resource, paramanayakam2025less, singh2024llm, yubeaton2025verithoughts}. For program synthesis and hardware designs, numerous methods and benchmarks have emerged for register-transfer level (RTL) code generation \citep{thakur2023benchmarking, liu2023verilogeval, pinckney2024revisiting, lu2024rtllm, yubeaton2025verithoughts, thakur2024verigen}, electronic design automation (EDA) tool scripting~\cite{wu2024chateda}, accelerator design~\cite{fu2023gpt4aigchip}, hardware design language (HDL) error debugging~\cite{hemadri2025veriloc, tsai2024rtlfixer}, and post-synthesis metric estimation~\cite{abdelatty2025metrex}. However, to mature into expert-level ``HDL Copilots,'' agentic solutions must navigate disparate EDA workflows, nuanced design trade-offs, and complex synthesis pipelines~\cite{thakur2024verigen, yubeaton2025verithoughts}.

Recent approaches fine-tune large language models (LLM) on curated RTL/HDL benchmarks as pairs of natural language prompts (RTL design questions) and their corresponding hardware implementations~\cite{thakur2023benchmarking, liu2023verilogeval, pinckney2024revisiting, yubeaton2025verithoughts, thakur2024verigen}. Though these methods demonstrate strong performance~\cite{liu2023verilogeval, yubeaton2025verithoughts}, they rely on costly fine-tuning~\cite{pinckney2024revisiting}, which requires substantial GPU budgets and might generalizes poorly to adjacent HDL tasks~\cite{abdelatty2025metrex}. On the other hand, frontier Large Reasoning Models (LRMs), such as OpenAI's o4~\cite{jaech2024openai}, achieve robust results on RTL coding benchmarks~\cite{yubeaton2025verithoughts} without fine-tuning, but shift the computational burden from training to inference~\cite{stamoulis2025geo}.

In this work, we draw inspiration from the novel paradigm of automated multi-agent workflow generation~\cite{zhang2025maas, niu2025flow, zhang2025aflow, hu2025adas}. Recent methods, such as MaAS~\cite{zhang2025maas}, Flow~\cite{niu2025flow}, and AFlow~\cite{zhang2025aflow}, improve task performance-cost trade-offs compared to monolithic LLM prompting with strong generalizability. However, as discussed in~\cite{zhang2025maas}, these methods primarily focus on question-answering (QA) with wiki-trivia and math quizzes and coding tasks. Consequently, a gap might persist between ``general knowledge'' domains, where simple prompting operators such as Debate~\cite{du2023debate} yield robust performance on multiple-choice QA queries, versus domain-specific RTL design problems~\cite{yubeaton2025verithoughts}.

Our \textbf{key insight} is to integrate HDL verification checks directly into the workflow generation process: we dynamically provide  agentic reasoning with design logs and error messages from RTL/EDA synthesis tools  to guide workflow creation. This simple yet effective solution enables us to experiment with agentic reasoning across various design objectives, such as optimizing for post-synthesis goals through PPA (power, performance, and area)–aware prompting. Across state-of-the-art benchmarks~\cite{yubeaton2025verithoughts, liu2023verilogeval}, our approach achieves accuracy on par or higher compared to prior methods by up to 7\% \textit{pass@k}, while requiring only a few hundred ``training'' examples for controller tuning, representing an \textbf{order-of-magnitude} less supervision than full fine-tuning. 

\begin{figure*}[t]
  \centering
  \includegraphics[width=\linewidth]{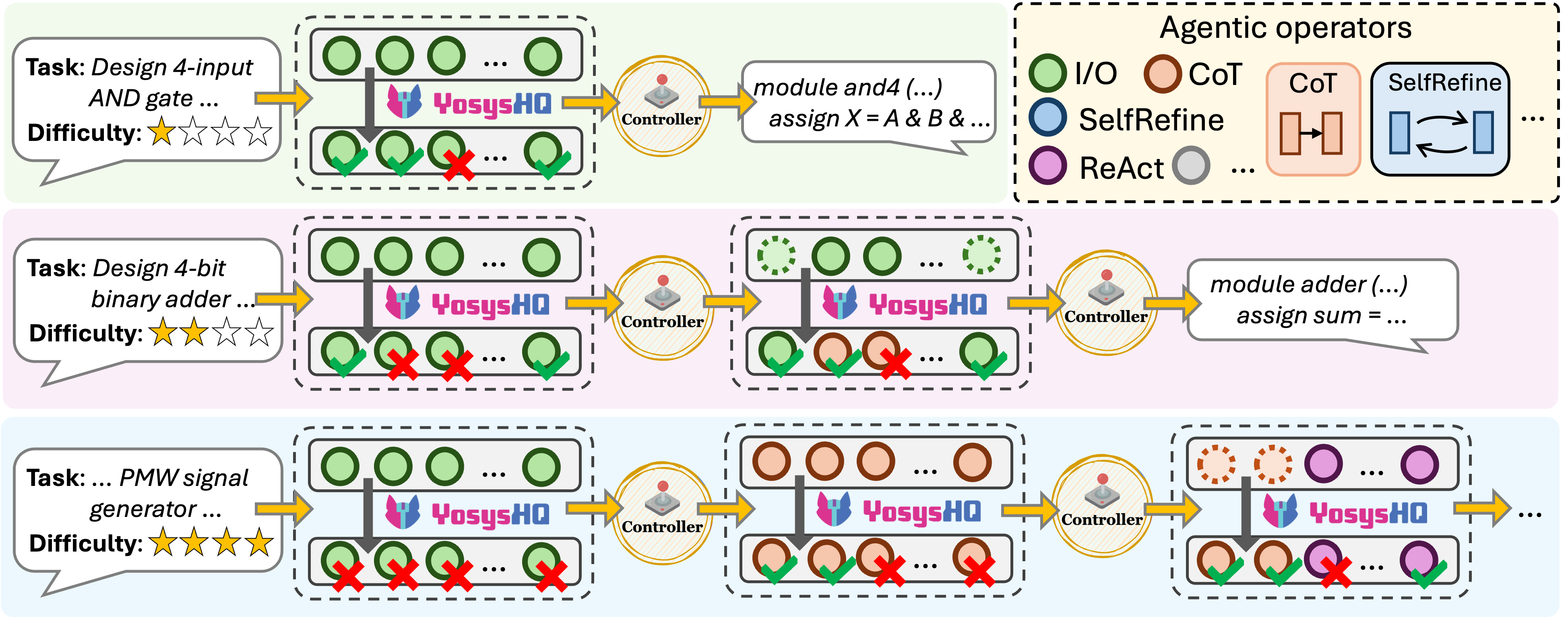}
  \caption{\textit{VeriMaAS}: Given RTL tasks with varying difficulty, we adaptively sample agentic operators: at each step, the selected operators are evaluated against formal verification and synthesis EDA tools. The workflow controller receives synthesis logs to dynamically refine operator selection.}
  \vspace{-10pt}
  \label{fig:method}
\end{figure*}

\section{Methodology}
\label{sec:method} 

Figure~\ref{fig:method} illustrates our approach. Given an RTL design task, \textit{VeriMaAS} adaptively samples a set of reasoning operators based on the input query and task difficulty. In each state, the candidate Verilog designs are executed via a synthesis and verification pipeline using Yosys~\cite{wolf2013yosys} and OpenSTA~\cite{ajayi2019toward}, and the resulting log-error messages are used as feedback to inform subsequent decision steps to refine the reasoning strategy or terminate the generation of code and return the current pool of samples.

\textbf{Solution Space}. We define solution space $\mathbb{O}$ as the set of available agentic operators: Zero-shot I/O, Chain-of-Thought (CoT)~\cite{wei2022cot}, ReAct~\cite{yao2023react}, Self-Refine~\cite{madaan2023selfrefine}, and Debate~\cite{du2023debate}. We denote the multi-agent solution in $\mathbb{O}$ as $\mathcal{O} = \{O_i \in \mathbb{O} \mid i = 1, \dots, K\}$. In fact, most existing prompting schemes can be viewed as single-solution operator sequences. For example, always using zero-shot CoT prompting corresponds to $\mathcal{O} = \{O_{\text{CoT}}\}$. Similarly, a Self-Refine setup that corrects a previous solution can be represented as a two-step sequence $\mathcal{O} = \{O_{\text{CoT}}, O_{\text{SelfRefine}}\}$. Given a user query, our goal is to find a composition of operators $\mathcal{O}$ that yields top RTL designs. For code generation, performance is typically assessed using \textit{pass@k} metrics~\cite{chen2021evaluating} over a fixed set of model outputs. As \textit{pass@k} is typically computed over $K=20$ samples, we set $|\mathcal{O}| = K = 20$.

\textbf{Agentic Controller}. The core of multi-agentic workflow automation lies in the \textit{controller} module $\mathcal{C}$ which dynamically selects the prompting operators for each task: $\mathcal{C}$ takes as input the query $q$, the operators $\mathbb{O}$, and the answers $\mathcal{A}_{\mathcal{O}_{current}}$ from the current solution $\mathcal{O}_{current}$, and outputs an updated solution $\mathcal{O}_{new}$. We employ a \textit{cascading controller}~\cite{zhang2025maas} that selects increasingly complex operators at each stage, following $\{O_{\text{I/O}} \rightarrow O_{\text{CoT}} \rightarrow O_{\text{ReAct}} \rightarrow O_{\text{SelfRefine}} \rightarrow O_{\text{Debate}}\}$. At each stage $c$, the controller $\mathcal{C}$ computes a confidence score $s_c$ indicating whether to proceed to the next operator in the cascade. If $s_c$ falls below a threshold $\tau_c$, the controller returns the current solution vector $\mathcal{A}_{\mathcal{O}_{c}}$; otherwise, it adds a new operator $O_{c+1}$ to the possible solution set and continues. We denote the set of per-stage thresholds as $\mathcal{T} = \{\tau_1, \tau_2, \dots, \tau_C\}$. 

In our implementation, we leverage the fact that formal verification and synthesis provide a strong signal of design complexity. Intuitively, if the majority of candidate solutions (\textit{e.g.}, initially generated using I/O) fail to compile or pass Yosys checks~\cite{wolf2013yosys}, this suggests that the task may require more sophisticated reasoning operators. At each cascade stage, we run $K = 20$ Verilog candidates $\mathcal{A}_{\mathcal{O}_{c}}$ through Yosys for synthesis and area estimation, and through OpenSTA~\cite{ajayi2019toward} for timing and power analysis. We compute $s_c = \texttt{Score}(\mathcal{A}_{\mathcal{O}_{c}})$ as the percentage of failing designs due to errors in verification, synthesis (area), runtime, or power analysis. If this percentage exceeds the stage-specific threshold $\tau_c \in [0,1]$, the controller proceeds to the next cascade stage.

\textbf{Problem formulation}. The controller $\mathcal{C}$ is parameterized by the per-stage thresholds $\mathcal{T} = \{\tau_1, \dots, \tau_C\}$. Let $D$ denote a dataset consisting of queries $q$ and their corresponding oracle Verilog solutions $a$. Our objective is to learn the thresholds $\mathcal{T}$ that generate high-quality solutions with minimal token cost~\cite{zhang2025maas}. Formulated as a multi-objective optimization problem, we write~\cite{zhang2025maas}:
\begin{equation}
\max_{\mathcal{T}}  \mathbb{E}_{(q,a) \sim D} \left[ U(\mathcal{T}; q, a, \mathbb{O}) - \lambda \cdot C(\mathcal{T}; q, a, \mathbb{O}) \right]
\label{eq:problem_formulation}
\end{equation}
Here, $U(\cdot)$ and $C(\cdot)$ denote the utility and cost of solving the subset of query-answer pairs $D$ under a threshold configuration $\mathcal{T}$ and operator set $\mathbb{O}$. We compute $U(\cdot)$ and $C(\cdot)$ as the \textit{pass@k} score and the average number of tokens per query, respectively. We set $\lambda=1e^{-3}$ as in~\cite{zhang2025maas}.

Following~\cite{zhang2025maas}, we randomly sample 500 datapoints from the VeriThoughts~\cite{yubeaton2025verithoughts} \textit{training} set. For each datapoint, we compute \textit{pass@k} and token cost against the ground-truth design. Based on synthesis results for $K=20$ candidates per query, we count how many fail to pass Yosys and OpenSTA checks. To determine thresholds $\mathcal{T}$, we aggregate failure counts across all samples and compute the 20th, 40th, 60th, and 80th percentiles (\textit{i.e.}, corresponding to our five operators in $\mathbb{O}$). We note that this ``tuning'' procedure requires a few hundred datapoints, representing an order-of-magnitude reduction in training cost compared to the tens of thousands of samples needed for full fine-tuning~\cite{yubeaton2025verithoughts}. 

\begin{table}
  \caption{\textit{Pass@k} \textit{vs}. Instruct LLMs, Reasoning LRMs, and fine-tuned RTL-coding models~\cite{shang2024rtlcoder, yubeaton2025verithoughts}.}
  \label{table:results}
  \centering
  \resizebox{1.0\textwidth}{!}{ 
  \begin{tabular}{llllll}
    \toprule
    \multirow{2}{*}{Model} &  \multirow{2}{*}{Method} & 
    \multicolumn{2}{c}{VeriThoughts~\cite{yubeaton2025verithoughts}} &  
    \multicolumn{2}{c}{VerilogEval~\cite{pinckney2024revisiting}} \\
    \cmidrule(r){3-4}\cmidrule(r){5-6}
    &  & \textit{Pass@1} & \textit{Pass@10} & \textit{Pass@1} & \textit{Pass@10} \\
    \midrule
    GPT 4o-mini                 & Instruct         &   80.64 & 90.87 & 50.26 & 61.02 \\
    GPT 4o-mini+\textbf{VeriMaAS}   & Agent workflow   &  83.09 \green{2.45} & 92.85 \green{1.98} & 52.05 \green{1.79} & 64.02 \green{3.00} \\
    \midrule
    o4-mini                     & Reasoning         &  93.85 & 97.88 & 75.67 & 85.13 \\
    o4-mini+\textbf{VeriMaAS}   & Agent workflow    &  94.09 \green{0.24} & 98.17 \green{0.29} & 76.15 \green{0.48} & 84.50 \red{0.63} \\
    \midrule
    Qwen2.5-7B  & Instruct    & 44.90 & 82.33 & 22.92 & 51.47 \\
    RTLCoder-7B~\cite{shang2024rtlcoder} & RTL Fine-tuned & -- & -- & 34.60 \green{11.68} & 45.50 \red{5.97} \\
    Qwen2.5-7B+\textbf{VeriMaAS}  & Agent workflow & 56.62 \green{11.72} & 86.29 \green{3.96} & 29.10 \green{6.18} & 56.45 \green{4.98} \\
    \midrule
    Qwen2.5-14B    & Instruct & 67.89 & 94.13 & 33.78 & 62.04 \\
    VeriThoughts-14B~\cite{yubeaton2025verithoughts} & RTL Fine-tuned & 78.50 \green{10.61} & 92.10 \red{2.03} & 43.70 \green{9.92} & 55.14 \red{6.90} \\
    Qwen2.5-14B+\textbf{VeriMaAS}   & Agent workflow & 74.24 \green{6.35}  &  95.78 \green{1.65} & 41.47 \green{7.69} & 62.48 \green{0.44} \\
    \midrule
    Qwen3-8B     & Reasoning & 84.11 & 98.82 & 58.21 & 74.64 \\
    RTLCoder-DeepSeek-7B~\cite{shang2024rtlcoder}  &  RTL Fine-tuned  & -- & -- & 39.70 & 51.90 \\
    Qwen3-8B+\textbf{VeriMaAS}  & Agent workflow & 88.13 \green{4.02} & 99.05 \green{0.23} & 59.87 \green{1.66} & 74.18 \red{0.46} \\
    \midrule
    DeepSeek-R1-Qwen-14B~\cite{guo2025deepseek} & Reasoning & 46.20 & 89.10 & 38.70 & 69.00 \\
    Qwen3-14B   & Reasoning  & 89.35 & 98.64 & 65.87 & 75.62 \\
    Qwen3-14B+\textbf{VeriMaAS} & Agent workflow  & 92.16 \green{2.81} & 98.75 \green{0.11} & 66.96 \green{1.09} & 75.71 \green{0.09} \\
    \bottomrule
  \end{tabular}
  }
\end{table}

\section{Results}
\label{sec:res}

\textbf{Experimental Setup}. We evaluate performance on two state-of-the-art benchmarks, VerilogEval~\cite{liu2023verilogeval} and VeriThoughts~\cite{yubeaton2025verithoughts}, and report \textit{pass@1} and \textit{pass@10} scores over $20$ samples following the VeriThoughts repo~\cite{yubeaton2025verithoughts}. We use Yosys~\cite{wolf2013yosys} for verification and area estimation, and OpenSTA~\cite{ajayi2019toward} for timing and static power analysis. All designs are synthesized using the Skywater 130nm PDK~\cite{skywater2020pdk}, following the MetRex synthesis benchmark~\cite{abdelatty2025metrex}. We evaluate proprietary and open-weight models, covering Instruct and Reasoning baselines, as well as existing fine-tuned RTL approaches~\cite{yubeaton2025verithoughts, hemadri2025veriloc}.

\begin{table}
  \caption{\textit{Pass@k} scores and token cost \textit{vs}. single-agent solvers. We report the relative change \textit{vs}. the Table~\ref{table:results} baselines. We \textbf{bold} the best results and \underline{underline} the runner-ups.}
  \label{table:tradeoff}
  \centering
  \resizebox{1.0\textwidth}{!}{ 
  \begin{tabular}{llllllll}
    \toprule
    \multirow{2}{*}{Model} &  \multirow{2}{*}{Prompting} & 
    \multicolumn{3}{c}{VeriThoughts~\cite{yubeaton2025verithoughts}} &  
    \multicolumn{3}{c}{VerilogEval~\cite{pinckney2024revisiting}} \\
    \cmidrule(r){3-4}\cmidrule(r){5-6}
    & & \textit{Pass@1} & \textit{Pass@10} & \textit{Tokens (k)} & \textit{Pass@1} & \textit{Pass@10} & \textit{Tokens (k)} \\
    \midrule
    \multirow{4}{*}{o4-mini} & + CoT~\cite{wei2022cot}  & \underline{94.11} \green{0.26} & 97.86 \red{0.02} & 1.10 \neutral{1.09$\times$} & \underline{76.06} \green{0.39} & \underline{84.35} \red{0.78} & 1.60 \neutral{1.06$\times$}\\
    & + ReAct~\cite{yao2023react}     & 91.96 \red{1.89} & 98.04 \green{0.16} & 1.70 \neutral{1.68$\times$} & 74.33 \red{1.34} & 84.10 \red{1.03} & 2.14 \neutral{1.42$\times$}\\
    & + SelfRefine~\cite{madaan2023selfrefine}     & \textbf{94.31} \green{0.46} & \textbf{98.57} \green{0.69} & 2.24 \neutral{2.22$\times$} & 75.71 \green{0.04} & 84.05 \red{1.08} & 3.23 \neutral{2.14$\times$}\\
    & +\textbf{VeriMaAS}  & 94.09 \green{0.24} & \underline{98.17} \green{0.29} & 1.21 \neutral{1.20$\times$} & \textbf{76.15} \green{0.48} & \textbf{84.50} \red{0.63}& 1.71 \neutral{1.13$\times$}\\
    \midrule
    \multirow{4}{*}{GPT 4o-mini} & + CoT~\cite{wei2022cot}   & 82.25 \green{1.61} & 92.05 \green{1.18} & 0.71 \neutral{1.42$\times$} & 51.25 \green{0.99} & 62.07 \green{1.05} & 0.77 \neutral{1.33$\times$}\\
    & +ReAct~\cite{yao2023react}   & 82.77 \green{2.13} & \textbf{93.10} \green{2.23} & 1.33 \neutral{2.66$\times$} & \textbf{54.81} \green{4.55} & \textbf{67.31} \green{6.29} & 1.36 \neutral{2.34$\times$}\\
    & +SelfRefine~\cite{madaan2023selfrefine}  & \underline{83.02} \green{2.38} & 92.48 \green{1.61} & 1.45 \neutral{2.90$\times$} & 51.47 \green{1.21} & 61.75 \green{0.73} & 1.59 \neutral{2.74$\times$}\\
    & +\textbf{VeriMaAS} & \textbf{83.09} \green{2.45} & \underline{92.85} \green{1.98} & 1.26 \neutral{2.52$\times$} & \underline{52.05} \green{1.79} & \underline{64.02} \green{3.00} & 0.85 \neutral{1.47$\times$}\\
    \midrule
    \multirow{4}{*}{Qwen2.5-14B} & + CoT~\cite{madaan2023selfrefine}   & 68.11 \green{0.22} & 95.09 \green{0.96} & 0.70 \neutral{1.08$\times$} & 37.56 \green{3.78} &  \textbf{65.04} \green{3.00} & 0.91 \neutral{1.07$\times$}\\
    & + ReAct~\cite{yao2023react}     & 62.53 \red{5.36} & 93.38 \red{0.75} & 1.09 \neutral{1.68$\times$} & 24.01 \red{9.77} & 57.57 \red{4.47} & 1.17 \neutral{1.38$\times$}\\
    & + SelfRefine~\cite{madaan2023selfrefine}  & \textbf{74.71} \green{6.82} & \textbf{95.96} \green{1.83} & 1.49 \neutral{2.29$\times$} & \underline{41.67} \green{7.89} & \underline{63.55} \green{1.51} & 1.85 \neutral{2.18$\times$}\\
    & +\textbf{VeriMaAS}  & \underline{74.24} \green{6.35}  & \underline{95.78} \green{1.65} & 1.21 \neutral{1.86$\times$} & 41.47 \green{7.69} & 62.48 \green{0.44} & 1.61 \neutral{1.89$\times$}\\
    \bottomrule
  \end{tabular}
  }
\end{table}

\begin{table}
  \caption{Post-synthesis delta with and without PPA-aware optimization. We report the relative change in PPA metrics; \textit{Pass@10} deltas are on entire benchmark \textit{vs}. +\textbf{VeriMaAS} baselines (Table~\ref{table:results}).}
  \label{table:ppa}
  \centering
  \resizebox{1.0\textwidth}{!}{ 
  \begin{tabular}{llrrrlrrr}
    \toprule
    \multirow{2}{*}{Model} & 
    \multicolumn{4}{c}{VeriThoughts\texttt{-PPA-Tiny}~\cite{yubeaton2025verithoughts}} &  
    \multicolumn{4}{c}{VerilogEval\texttt{-PPA-Tiny}~\cite{pinckney2024revisiting}} \\
    \cmidrule(r){2-5}\cmidrule(r){6-9}
    & \textit{Pass@10} & \textit{$\Delta$Area\%} & \textit{$\Delta$Power\%} & \textit{$\Delta$Delay\%} 
    & \textit{Pass@10} & \textit{$\Delta$Area\%} & \textit{$\Delta$Power\%} & \textit{$\Delta$Delay\%} \\
    \midrule
    GPT-4o-mini    & 92.46 \red{0.39} & \ppagreen{9.18} & \ppared{1.6} & \ppagreen{10.32} & 62.93 \red{1.09} & \ppagreen{18.83} & \ppagreen{3.26} & \ppagreen{19.47} \\
    o4-mini      & 98.06 \red{0.11} & \ppagreen{14.86} & \ppagreen{0.00} & \ppagreen{15.87} & 84.18 \red{0.32} & \ppagreen{12.22} & \ppared{1.70} & \ppagreen{3.52} \\
    Qwen2.5-7B    & 86.33 \green{0.04} & \ppagreen{13.44} & \ppagreen{8.67} & \ppagreen{13.91} & 56.45 \green{0.00} & \ppagreen{28.79} & \ppared{4.07} & \ppagreen{24.58} \\
    Qwen2.5-14B   & 95.72 \red{0.06} & \ppagreen{16.8} & \ppagreen{14.57} & \ppagreen{21.39} & 62.33 \red{0.15} & \ppagreen{16.17} & \ppared{5.22} & \ppagreen{15.53} \\
    Qwen3-8B     & 99.04 \red{0.01} & \ppagreen{22.81} & \ppagreen{3.68} & \ppagreen{20.14} & 74.06 \red{0.12} & \ppagreen{9.98} & \ppagreen{6.04} & \ppagreen{9.03} \\
    Qwen3-14B    & 98.75 \green{0.00} & \ppagreen{9.99} & \ppared{2.12} & \ppagreen{9.94} & 75.64 \red{0.07} & \ppagreen{11.66} & \ppagreen{7.85} & \ppagreen{11.39} \\
    \bottomrule
  \end{tabular}
  }
\end{table}

\textbf{Main Results}.
Across both VeriThoughts and VerilogEval, VeriMaAS improves synthesis accuracy over strong single-agent prompting strategies and fine-tuned RTL models. On open-source LLMs, our framework yields up to 7–12\% gains in pass@1 compared to existing fine-tuned baselines. \textit{VeriMaAS} also consistently improves \textit{pass@10} across both benchmarks on top of base LLMs, indicating that the framework not only raises top-1 accuracy but also expands the pool of valid candidate designs.

Closed-source models see smaller but consistent gains, reflecting that multi-agent orchestration adds value even when base performance is already high. As summarized in Table 2, \textit{VeriMaAS} achieves these improvements with moderate token overhead, staying close to lightweight CoT~\cite{wei2022cot} prompting and below iterative strategies like Self-Refine~\cite{madaan2023selfrefine}. Table 2 also demonstrates that these gains hold for both \textit{pass@1} and \textit{pass@10} (with the exception of VerilogEval \textit{pass@10} which sees marginal drop across all operators).

\textbf{PPA-Aware Optimization}
Unlike fine-tuning that entangles performance objectives into model weights, our controller can be flexibly re-optimized for different goals. As a proof of concept, we set the cost term (Eq.~\ref{eq:problem_formulation}) to the Yosys-reported area $C= \textbf{Area}(\mathcal{T}; q, a, \mathbb{O})$.
We note that some benchmark tasks (\textit{e.g.}, NAND gates) might offer little room for PPA optimization, so we use OpenAI o4 as pseudo-oracle to recommend the top 20 designs where RTL changes are more likely to affect downstream PPA metrics. We refer to these subsets as \texttt{-PPA-Tiny}. Table~\ref{table:ppa} reports the deltas in area, power, and delay between the standard and PPA-optimized \textit{VeriMaAS} solutions. As expected, we observe area and runtime reductions by up to 28.79\%. On the other hand, power and \textit{pass@10} deltas suggest a trade-off: while many models show improvements, some (especially for VerilogEval) see slight power increases or marginal \textit{pass@10} decrease.

\section{Conclusion and Future Work}

We introduced VeriMaAS, a multi-agent framework that integrates formal verification feedback into RTL code generation, alleviating reliance on costly fine-tuning and long reasoning traces. To support broader collaboration and development, we maintain our prototype as an open-source WiP repository~\footnote{Link: \href{https://github.com/dstamoulis/maas/tree/verimaas/verithoughts}{\textcolor{magenta}{https://github.com/dstamoulis/maas/tree/verimaas/verithoughts}}: documentation-migration in progress}. Looking ahead, our aim is to further enhance the controller formulation by incorporating tree-search or RL-based policies following the latest workflow automation techniques from the broader AI community~\cite{hu2025adas, zhang2025aflow, niu2025flow}. Moreover, we motivate expanding our orchestration signals to commercial EDA tools, and integrating (commercial) PDKs towards a comprehensive synthesis and PPA optimization beyond the current proof-of-concept. 

\bibliographystyle{plainnat}
\bibliography{verimass_ml4sys}

\end{document}